\newif\ifabstract
\abstracttrue
\abstractfalse 
\newif\iffull
\ifabstract \fullfalse \else \fulltrue \fi

\documentclass[11pt]{article}
\usepackage{amsfonts}
\usepackage{amssymb}
\usepackage{amstext}
\usepackage[ruled,vlined]{algorithm2e}
\usepackage{amsmath}
\usepackage{xspace}
\usepackage{theorem}
\usepackage{graphicx}
\usepackage{url}
\usepackage{graphics}
\usepackage{colordvi}
\usepackage{colordvi}
\usepackage{subfigure}
\usepackage{lscape}
\usepackage[english]{babel}
\usepackage{booktabs}
\usepackage{tabularx}

\advance \oddsidemargin by -0.8in
\newcommand{\myparskip}{3pt}
\parskip \myparskip

{\hspace*{\fill}$\Box$\par\vspace{4mm}}

\begin{document}

\title{Impact of Distributed Routing of Intelligent Vehicles on Urban Traffic\footnote{\textbf{Published in the proceedings of IEEE International Smart Cities Conference 2018}}}
\author{Lama Alfaseeh \thanks{Email: {lama.alfaseeh@ryerson.ca}} \and Shadi Djavadian \thanks{Email: shadi.djavadian@ryerson.ca} \and Bilal Farooq\thanks{Laboratory of Innovations in Transportation (LiTrans), Ryerson, Canada, Email: {bilal.farooq@ryerson.ca}}}

\begin{titlepage}
	\maketitle
	
	\thispagestyle{empty}
	
\begin{abstract}
The impact of distributed dynamic routing with different market penetration rates (MPRs) of connected autonomous vehicles (CAVs) and congestion levels has been investigated on urban streets. Downtown Toronto network is studied in an agent-based traffic simulation. The higher the MPRs of CAVs--especially in the case of highly congested urban networks--the higher the average speed, the lower the mean travel time, and the higher the throughput.
\end{abstract}
	
\end{titlepage}

\section{Introduction}
Urban congestion has not only affected the traffic safety, air pollution, and the use of energy, but also its effects have been seen in the form of annual cost to commuters and to the economy. In 2006, a study by Metrolinx estimated that in 2031 cost to commuters and economy will sky rocket and reach \$7.8, and \$7.2 respectively in the Greater Toronto and Hamilton Area \cite{r1}. 
It has been shown that the higher the efficiency of a routing system, the less the congestion and better the network performance and smaller travel time \cite{r2}. In literature, distributed routing systems were shown to overcome many of the centralized routing systems shortcomings that are mainly: the massive capital investment required to build a centralized system, high sensitivity to system failures, complexity of system upgrades, and lack of relativeness of information to a specific trip in general \cite{r2}. Furthermore, it has been shown that when dealing with a high-level of complexity in large-scale traffic networks, distributed outperforms central routing systems, as local data is used in order to take quick control actions when required due to perturbations in the traffic network \cite{r3}. 

Taking advantage of the Information and Communication Technology (ICT) development, intelligent vehicles are employed to enhance the traffic characteristics \cite{r4}. Many studies have been conducted and most of them used hypothetical highway segment. Olia et al. \cite{r5} illustrated the impact on the highway traffic capacity in the case of connected autonomous vahicles (CAVs), Autonomous Vehicles (AVs), and Human Driven Vehicles (HDVs) for different combinations. Several other studies have shown in highway settings that the higher the Market Penetration Rate (MPR) of intelligent vehicles the better the traffic network characteristics \cite{r5,r6}.

Moreover, integrating vehicle routing with the intelligent vehicles technology has a substantial contribution, it was found that CAVs have impressive potentials in overcoming the drawbacks of self-organization when choosing the route in which drivers are no longer decision makers when employing CAVs and compliance to the route guidance system is 100\%. It was shown that when CAVs are used, traffic conditions are improved \cite{r7}. Recently, Djavadian and Farooq \cite{r7} proposed an End-to-End dynamic distributed routing system (E2ECAV) based on a network of intelligent intersections and level 5 CAVs where full cooperation and coordination is expected. The results from their case study of downtown Toronto for 100\% CAV fleet showed considerable increase in throughput and decrease in average travel time. 
Our study is the first of its kind in which the impact of different MPRs of CAVs is illustrated on a large urban traffic network for the suggested distributed routing system \cite{r7}. Here, we adapted E2ECAV to systematically analyze the impact of congestion levels, Market Penetration Rates (MPRs) and E2ECAV parameters on the throughput, travel time, vehicle kilometers travelled, and other traffic flow indicators in downtown Toronto. 

\section{Background}
In the last two decades distributed routing systems have received considerable attention as they have shown promise in mitigating traffic congestion \cite{r8}. In reality, it is not always possible to have an accurate and well-founded origin-destination O-D matrix. It was proven that distributed routing systems can take over and deal with the lack of reliable O-D matrices. In distributed routing systems detectors are scattered in the network and have partial information that gets to be processed and provide the vehicles with guidance toward their destinations \cite{r3}.  Hawas and Mahmassani \cite{r3} compared the robustness of the two types of routing systems in the case of different traffic incidents e.g. lane blockage and for different levels and durations. It was found that distributed routing systems caused less reduction in the capacity when compared to the centralized routing systems. The reason behind it was that the distributed route guidance was reactive whereas the centralized control system still used the same route guidance as it would for incident free condition.  Hence, great potentials of the distributed routing systems arise when employing CAVs in the context of traffic assignment \cite{r3}. 

In addition, Yang and Recker\cite{r2} have studied the effect of inter-vehicles communication feature for the suggested distributed routing system. They considered the MPR of 1\% and 2\% of the equipped vehicles, and different traffic conditions: highly congested, congested, and not congested. They found that travel time was decreased when the MPR was greater than 2\% of vehicles equipped \cite{r2}. Lee and Park\cite{r9} evaluated the efficiency of a route guidance strategies based on vehicle-infrastructure integration (VII). Their study illustrated a substantial reduction in network travel time with the presence of the route guidance system suggested. They considered the MPR of the vehicles equipped with VII of 5\%, 30\%, 70\%, and 100\% in which they found that the higher the MPR of equipped vehicles the better the network characteristics. In addition, they took different congestion levels: normal, congested, and highly congested. Finally, updating interval, and driver compliance were considered as parameters \cite{r9}. Other studies of the distributed routing systems that took different traffic conditions and considered the total travel time in a network as an indicator can be found in \cite{r10, r11, r12}.

As per the aforementioned studies traffic condition, MPR and updating interval have significant impact on the performance of any distributed routing system that can be reflected directly on the traffic network performance. Many indicators were used in the literature to assess a traffic network performance in general e.g. throughput, travel time, speed, etc. Talebpour and Mahmassani took throughput, scatter in fundamental diagram of traffic flow as indicators to reflect the effect of using connected vehicles (CVs), and autonomous vehicles (AVs) \cite{r6}. In other studies, average speed \cite{r4}, average travel time, space average speed \cite{r13}, standard deviation of travel time \cite{r14}, and Vehicle Kilometers Traveled (VKT) \cite{r15} were the indicators. As distributed routing systems have been considered as a powerful tool for alleviating congestion. In the presence of intelligent vehicles, our study is first of its kind to examine the impact of different MPRs of CAVs in a distributed routing system for downtown Toronto network.

\section{Methodology}
Drivers tend to choose the shortest path to their destination without considering the effect of their choice on the system as a whole. Djavadian and Farooq \cite{r7} proposed an end-to-end distributed routing system (E2ECAV) based on the network of intelligent intersections and level 5 CAVs where drivers are no longer decision makers. 
In the E2ECAV proposed system, when CAVs arrive at an intersection they declare their destinations and they get to be routed based on the up to date routing table generated at the intersection. The routing table is based on the information provided by downstream intersections. The intersection routes the CAVs to the next intersection on their path in such a way that it minimizes individual CAVs as well as network travel time and maximizes capacity. On the other hand, HDVs routing strategy is based on the dynamic pre-trip shortest path. In other words, at the time HDVs enter the traffic network, the shortest path will be calculated based on the prevailed traffic condition and will not be changed afterwards. We assume that the efficiency of the proposed E2ECAV distributed routing system depends on the market penetration rate (MPR) of CAVs and the frequency of updating the information related to the network condition. In this paper, the effects of these parameters will be systematically investigated \cite{r7}. 
Before we illustrate the effect of different MPRs of CAVs, we developed the optimal configuration for E2EVAC parameters. In particular, the optimal updating interval, Intelligent Driver Model (IDM) \cite{r16} parameter set, and number of the shortest paths are determined based on the traffic characteristics in terms of throughput, total travel time and mean travel time. Different updating intervals studied are: 4sec, 30sec, 60sec, and 180sec. Two sets of IDM parameters for the CAVs have been investigated. The first set, regular IDM set, is similar to the HDVs (safe spacing distance= 2m, reaction time= 2sec). While the other CAV IDM set, reduced IDM set, is taken to be of half value of the regular IDM set for both safe spacing distance and reaction time. The rationale behind this is that these intelligent vehicles can perceive the changes in the environment quicker than humans and can maintain the acceleration and deceleration more smoothly. Acceleration and deceleration are the function of the engine technology and vehicle weight, so in our study, they are the same for both vehicle types. Moreover, number of the shortest paths (K-paths) are needed by E2ECAV. This parameter K-paths was set to 1, and 2.  Unlike Djavadian and Farooq \cite{r7}  study, that did not evaluate the MPR of CAVs nor different demand levels. This study evaluates various MPR of CAVs: (0\%, 5\%, 30\%, 50\%, 70\%, and 100\%) and three traffic conditions: highly congested, congested, and uncongested. It is profound to note that employing intelligent vehicles could affect the demand level. Nevertheless, this study is a supply analysis paper. In other words, the induced demand is not a considered aspect in this paper.

\section{Case Study}
Unlike other related studies that predominantly considered a single segment of a highway, an urban network of central downtown Toronto is used for our case study. Our network consists of 223 links, 76 nodes (intersections), and 26 centroids that are matched to the closest intersections. The morning peak hour from 7:45am to 8:00am is considered for this study.

\section{Discussion and Results}

\subsection{Updating Intervals for E2ECAV}
\begin{figure}[!t]
	\centering
	\hspace{0.35cm}
	\includegraphics[width=3in]{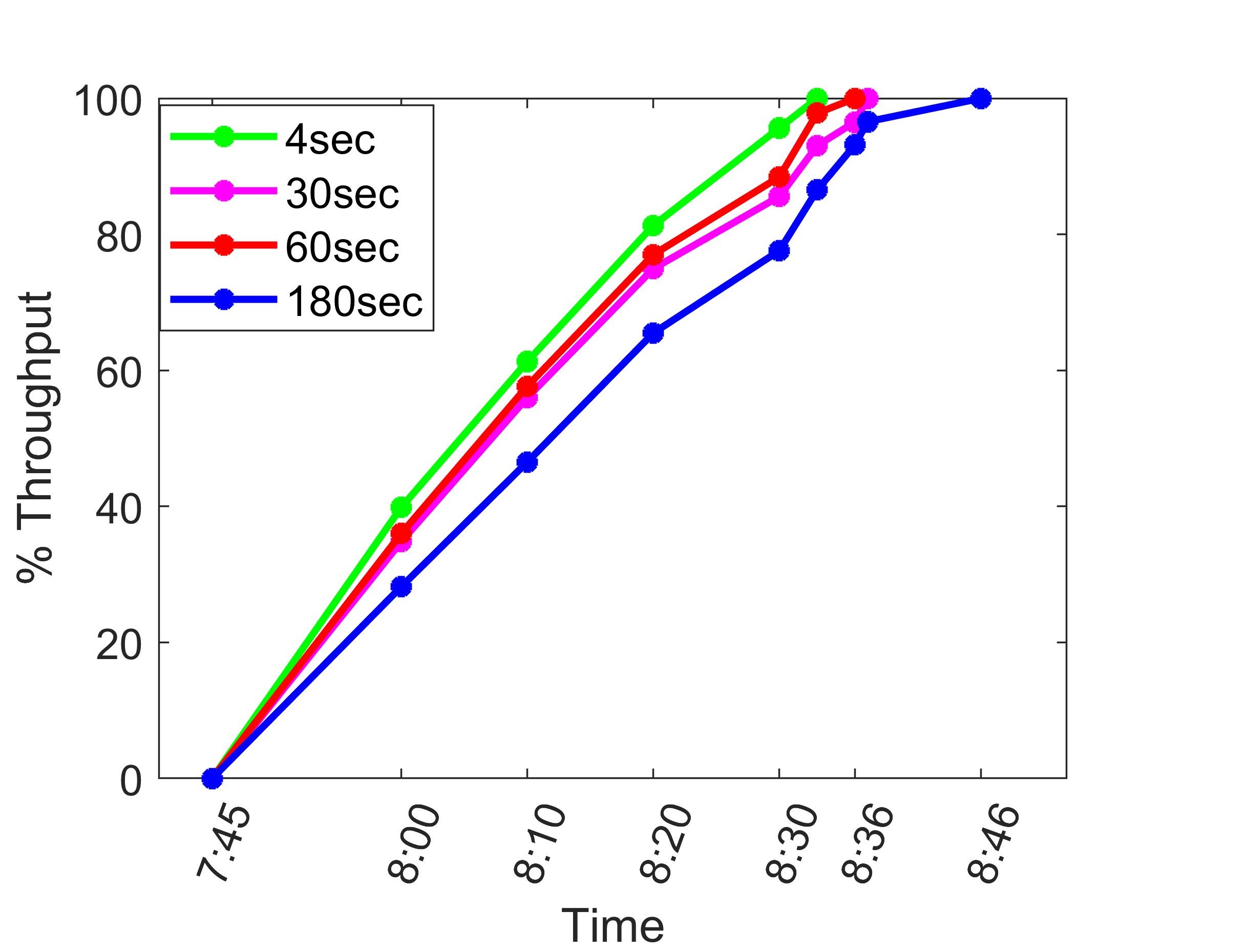}
	\caption{Throughput for Updating Intervals}
	\label{Throughput_Updating_intervals_Impact_Paper}
\end{figure}

\begin{figure}[!t]
	\centering
	\includegraphics[width=3in]{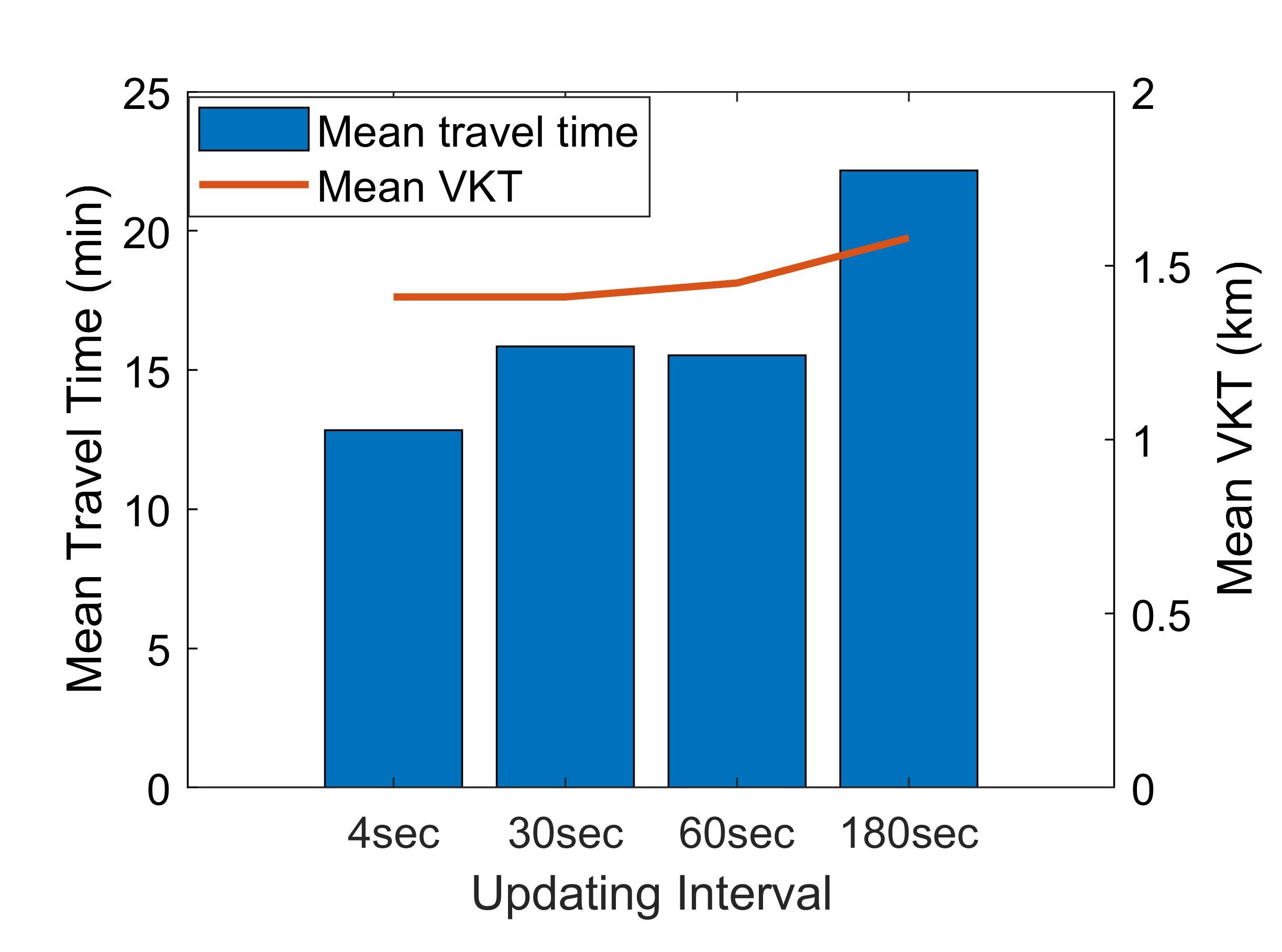}
	\caption{Mean Travel Time and Mean VKT for Updating Intervals}
	\label{MeanTT_VKT_updating_intervals_Impact_paper}
\end{figure}

Although in terms of the throughput, mean travel time, and travelled distance 4sec updating interval is the best as in Fig. \ref{Throughput_Updating_intervals_Impact_Paper} and Fig. \ref{MeanTT_VKT_updating_intervals_Impact_paper} respectively, 60sec is chosen because in real implementation communication delays and processing times may exceed 4sec and we also need to account for any communication errors requiring retransmission as well as detecting and recovering from downtimes. Moreover, the difference between the performance of 4sec and 60sec is only marginal. Finally and not to forget the high computational power required for the 4sec updating interval compared to the 60sec updating interval.

\subsection{IDM Parameters}
\begin{figure}[!t]
	\centering
	\includegraphics[width=3in]{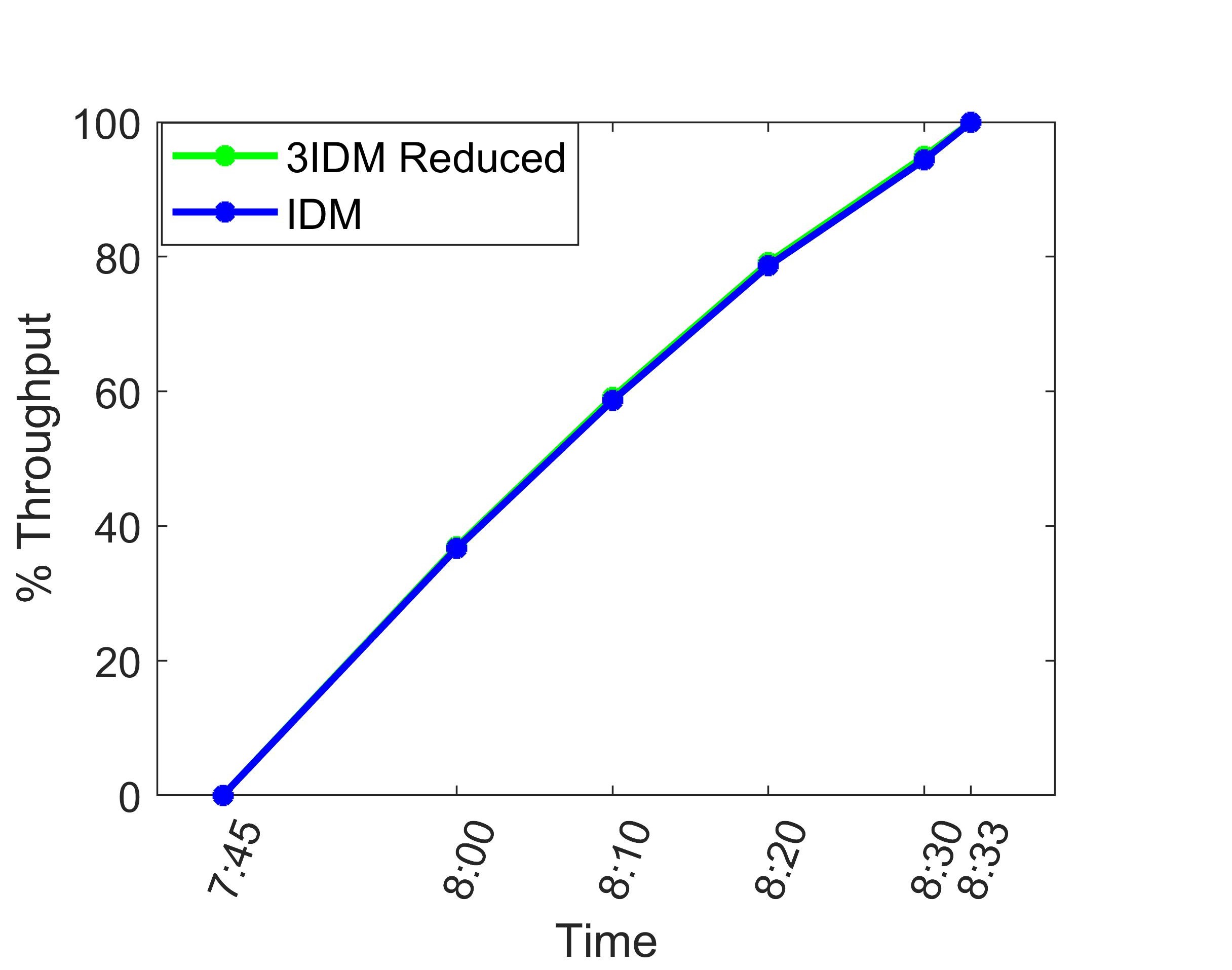}
	\caption{Throughput for IDM sets}
	\label{Throughput_IDM_Impact_Paper}
\end{figure}

\begin{figure}[!t]
	\centering
	\includegraphics[width=3in]{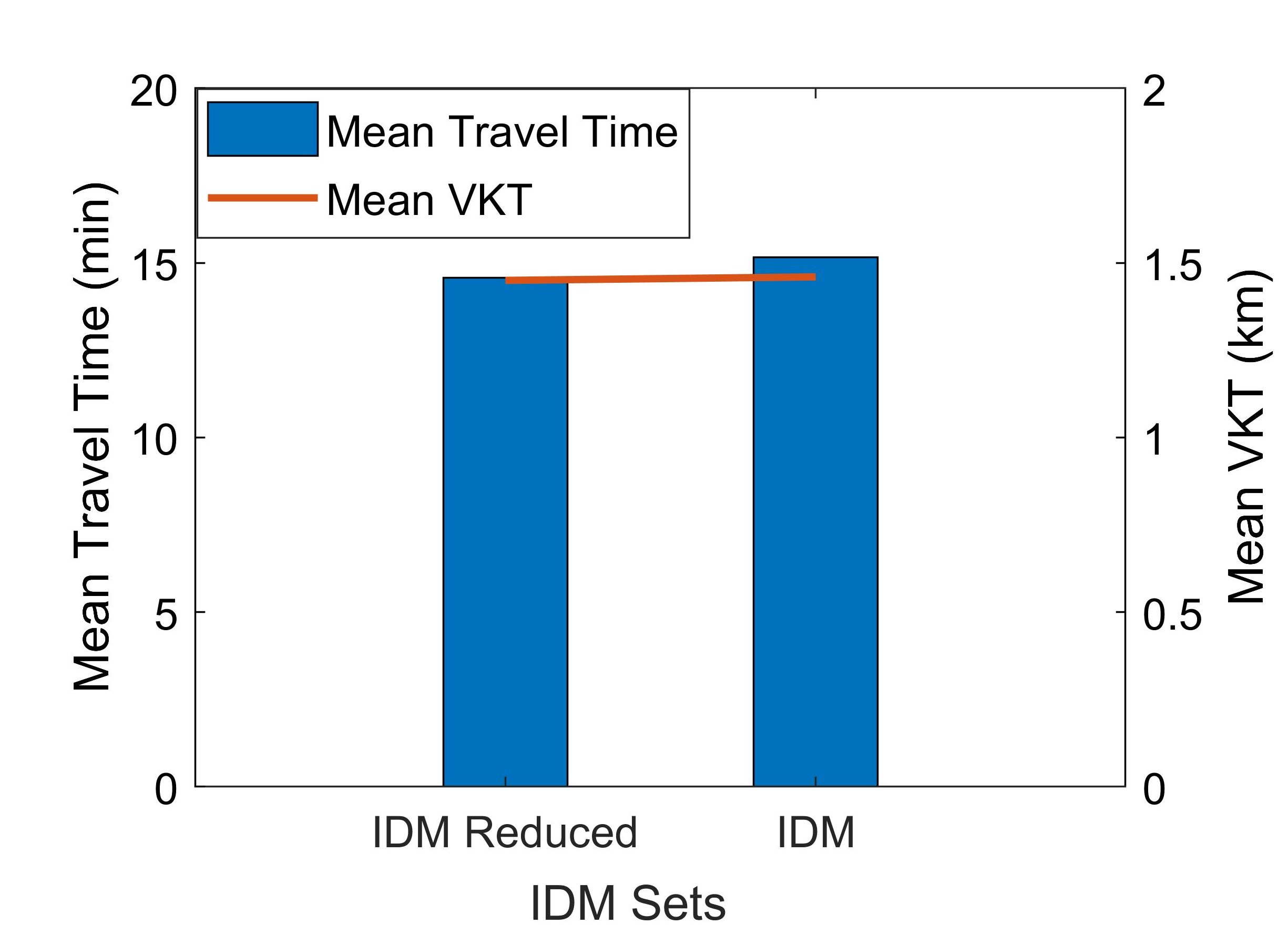}
	\caption{Mean Travel Time and Mean VKT for IDM Sets}
	\label{MeanTT_VKT_IDM_impact_paper}
\end{figure}

As illustrated in Table \ref{table1} we set the IDM parameters (safe spacing distance and reaction time) for HDV as fixed and studied the impact of two IDM sets of the CAVs. It can be observed in Fig. \ref{Throughput_IDM_Impact_Paper} and Fig. \ref{MeanTT_VKT_IDM_impact_paper} that reduced IDM parameters for CAVs resulted in a slightly better performance. This is due to the fact that the changes in safe spacing distance and reaction time will have less prominent effects in an urban area like downtown Toronto than on a highway. The two main reasons for the limited impact of the reduced IDM parameters, safe spacing distance and reaction time, in an urban area are: the links short length and the high frequency of the stop and go phenomena. Hence, employing reduced IDM parameters is good but not as good as employing connectivity in our case. Connectivity of the automated vehicle promises a profound effect in terms of urban traffic network improvement, as it can be used to direct the automated vehicles to the best routes based on the up-to-date information on the traffic conditions downstream. 
\begin{figure}[!t]
	\centering
	\includegraphics[width=3in]{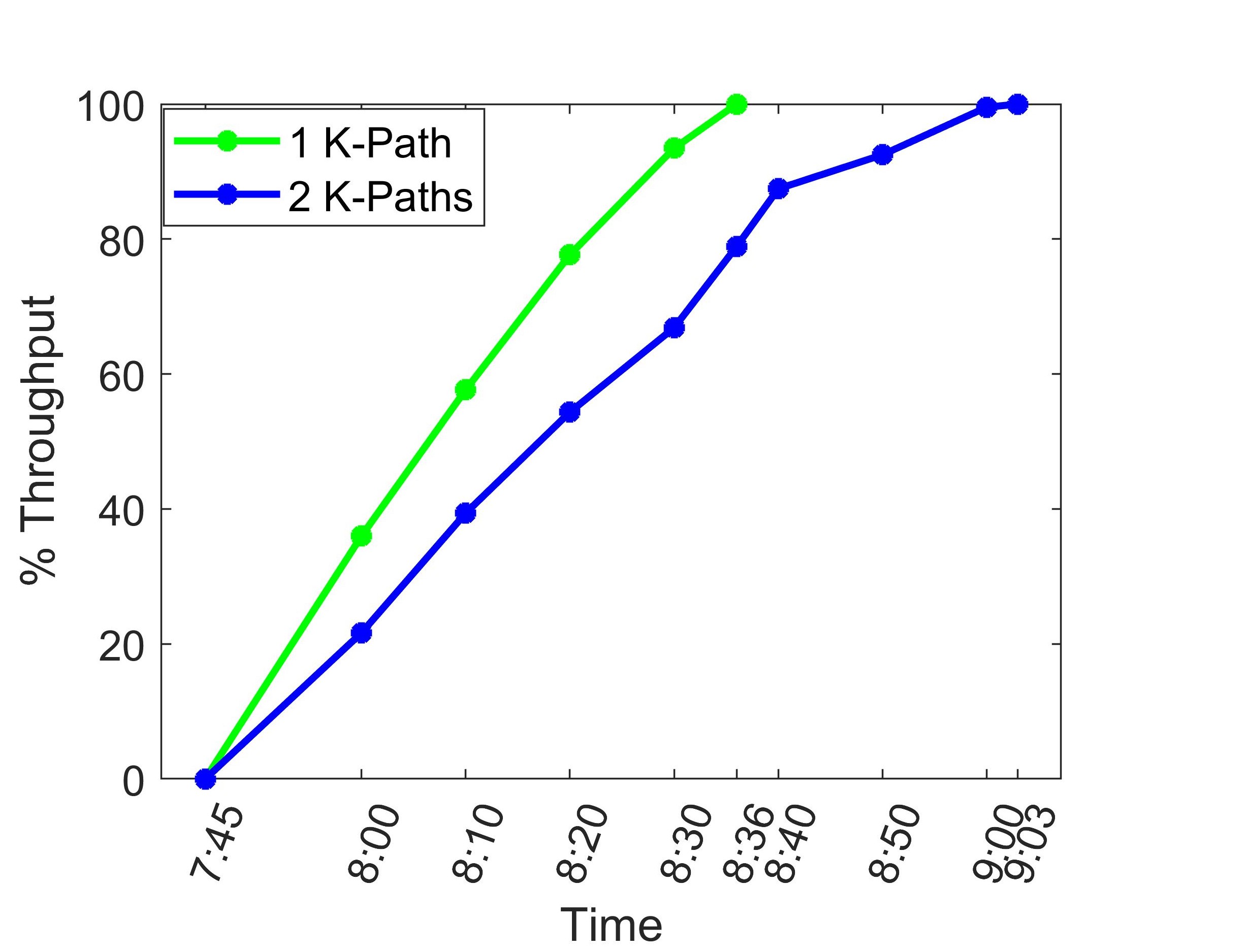}
	\caption{Throughput for K-paths}
	\label{Throughput_Kpath_Impact_Paper}
\end{figure}

\begin{figure}[!t]
	\centering
	\includegraphics[width=3in]{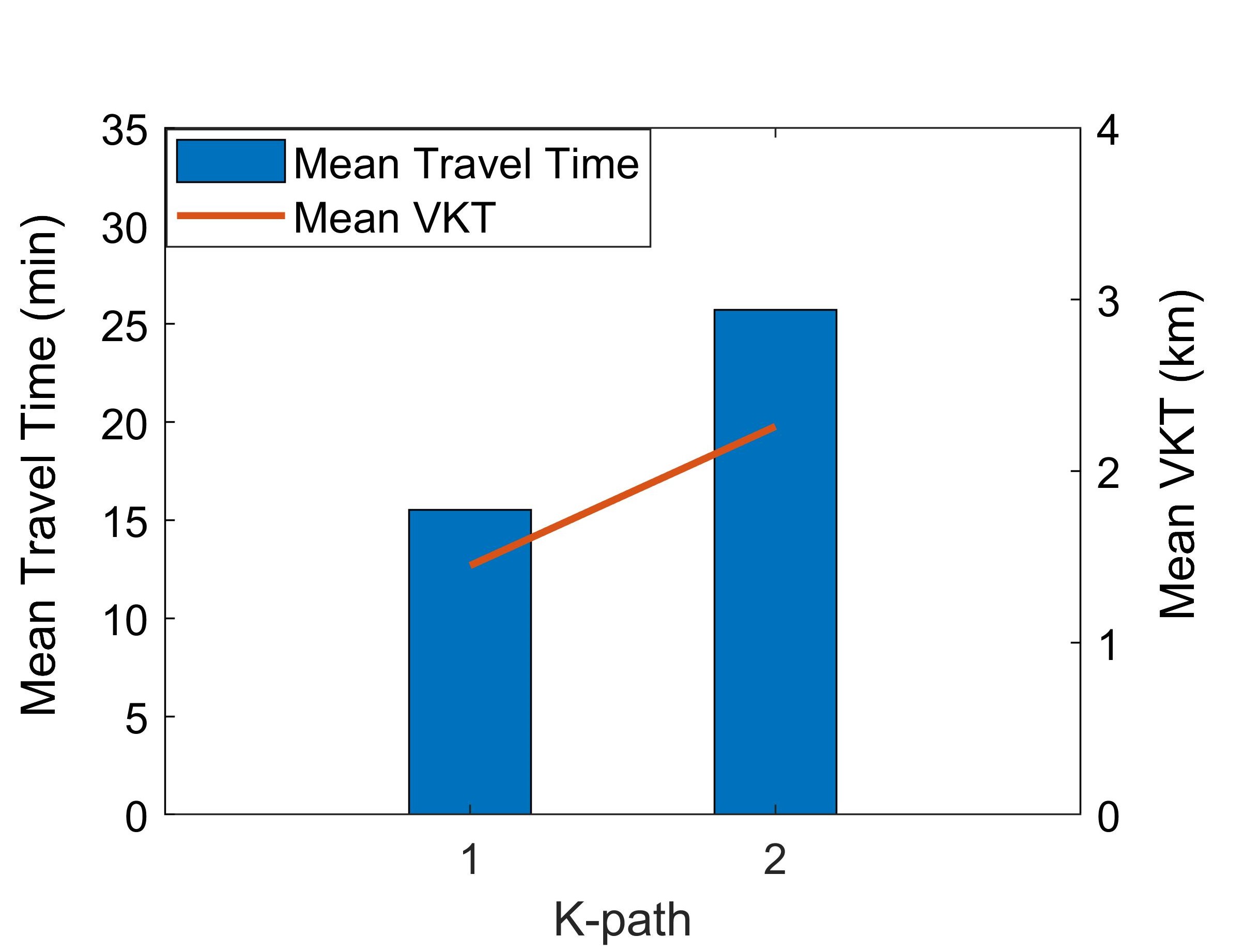}
	\caption{Mean Travel Time and Mean VKT for K-paths}
	\label{MeanTT_VKT_Kpath_impact_paper}
\end{figure}
\subsection{Number of Shortest Paths (K-paths) for E2ECAV}
Finally, the optimal number of shortest route in our study is K-path = 1 as shown in Fig. \ref{Throughput_Kpath_Impact_Paper} and Fig. \ref{MeanTT_VKT_Kpath_impact_paper}. It is important to note that when more shortest paths are used, vehicles may get to be rerouted to suboptimal shortest path with higher level of congestion more often and they also may end up in loops. This will result in the increase in travel times as well as distance travelled.
\begin{table}
	\caption{IDM parameters for HDVs and CAVs}
	\label{table1}
	\begin{center}
		\begin{tabular}[!t]{l c c}
			\hline
			IDM & Safe & Reaction\\
			& spacing (m) & time (sec)\\
			\hline
			\hline
			HDV & 2 & 2\\
			CAV & 2 & 2\\
			Reduced for CAV & 1 & 1\\
			\hline
			
		\end{tabular}
	\end{center}
\end{table}

\subsection{MPRs of CAVs and Different Traffic Conditions}
Once we have the optimal parameters of the updating interval, IDM set, and K-path, we proceed to analyze the effect of different MPRs of CAVs on the traffic network characteristics. We have taken three different traffic conditions: highly congested, congested, and uncongested. In addition, different MPRs of CAVs are considered: 0\%, 5\%, 30\%, 50\%, 70\%, 100\%. The impact of all the combinations of traffic conditions and MPRs on the traffic network based on several indicators is analyzed. 

\subsubsection{Mean Travel Time and Mean Distance Travelled} 
\begin{figure}[!t]
	\centering
	\includegraphics[width=3in]{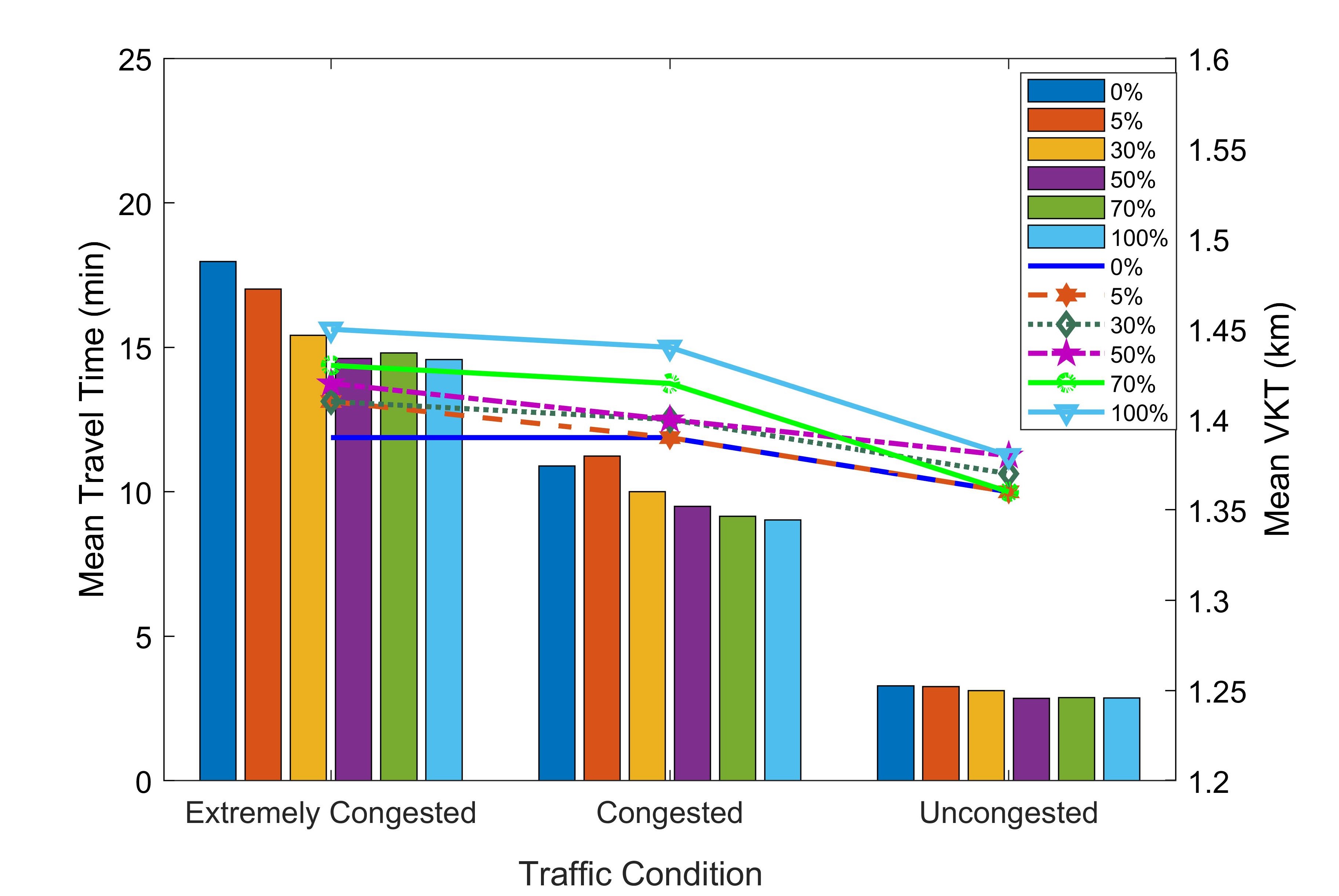}
	\caption{Mean Travel Time and Mean VKT for Traffic Conditions}
	\label{MeanTT_VKT_MPRs_DFs_Impact_Paper}
\end{figure}

It can be observed in Fig. \ref{MeanTT_VKT_MPRs_DFs_Impact_Paper} that in the three traffic conditions, the higher the MPR of CAVs the less mean travel time and more distance travelled. However, in the uncongested traffic network the difference of mean travel time and distance travelled is negligible. The enhancement in the mean travel time is due to the up to date information about the best route based on the traffic network condition. Moreover, intelligent intersections contributed by distributing the vehicles evenly on the network. On the other hand, the distance travelled increased when employing higher MPRs of CAVs. The increase is because intelligent intersections spread vehicles in the network and this caused longer travelled distance and re-routing might have contributed to the longer distance travelled. We can see the effect more pronounced in the highly congested and congested conditions. For the highly congested traffic network, employing 50\%, 70\%, and 100\% has relatively similar amount of reduction of the mean travel time. Around 18\% reduction of the mean travel time is observed when employing 100\% CAVs. Furthermore, for the highly congested traffic network a slight increase of the mean vehicle kilometers travelled and of 4\% when employing 100\% CAVs.

\subsubsection{Throughput} 
\begin{figure}[!t]
	\centering
	\includegraphics[width=3in]{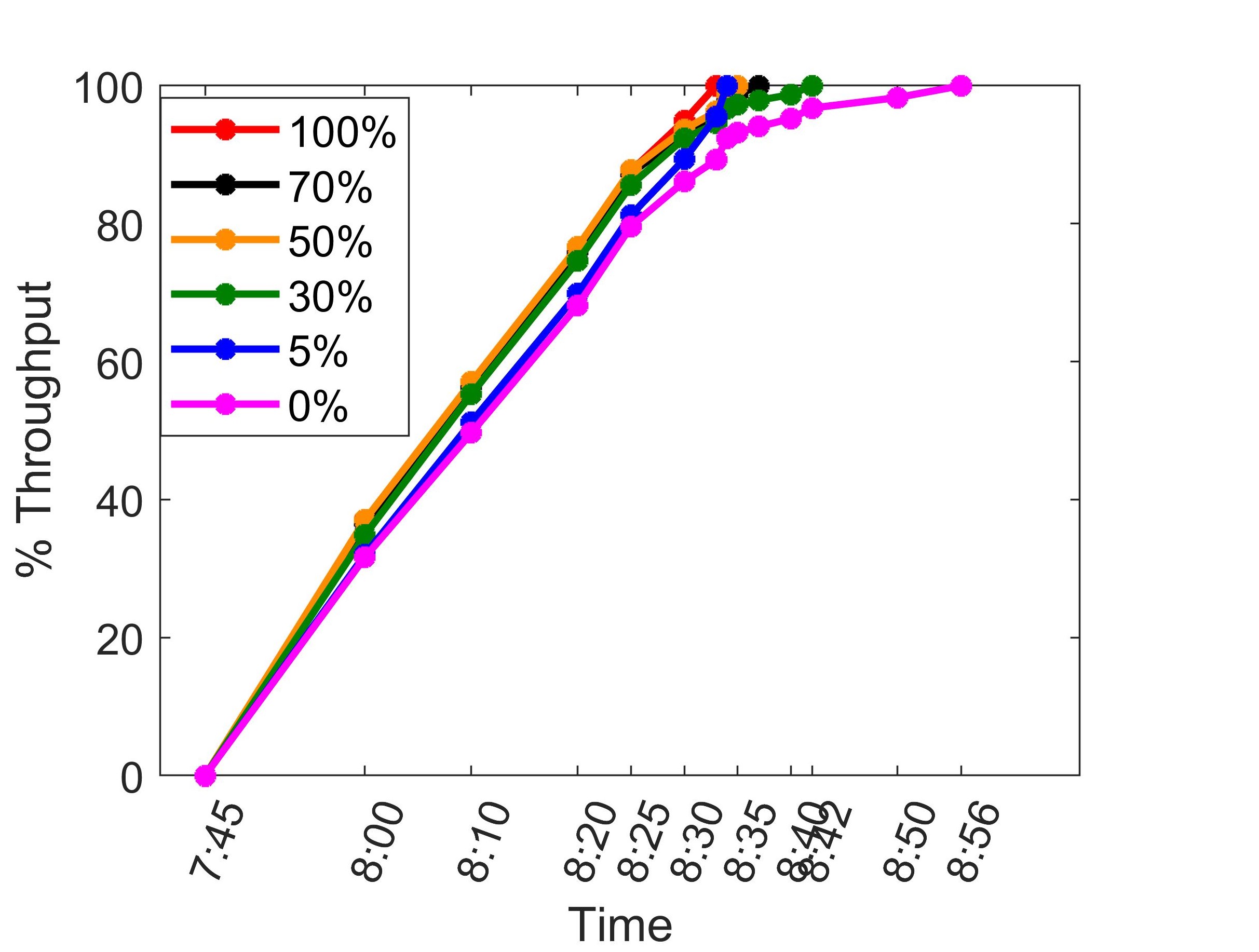}
	\caption{Throughput of Highly Congested Network for Different MPRs}
	\label{Throughput_HC_Impact_paper}
\end{figure}

\begin{figure}[!t]
	\centering
	\includegraphics[width=3in]{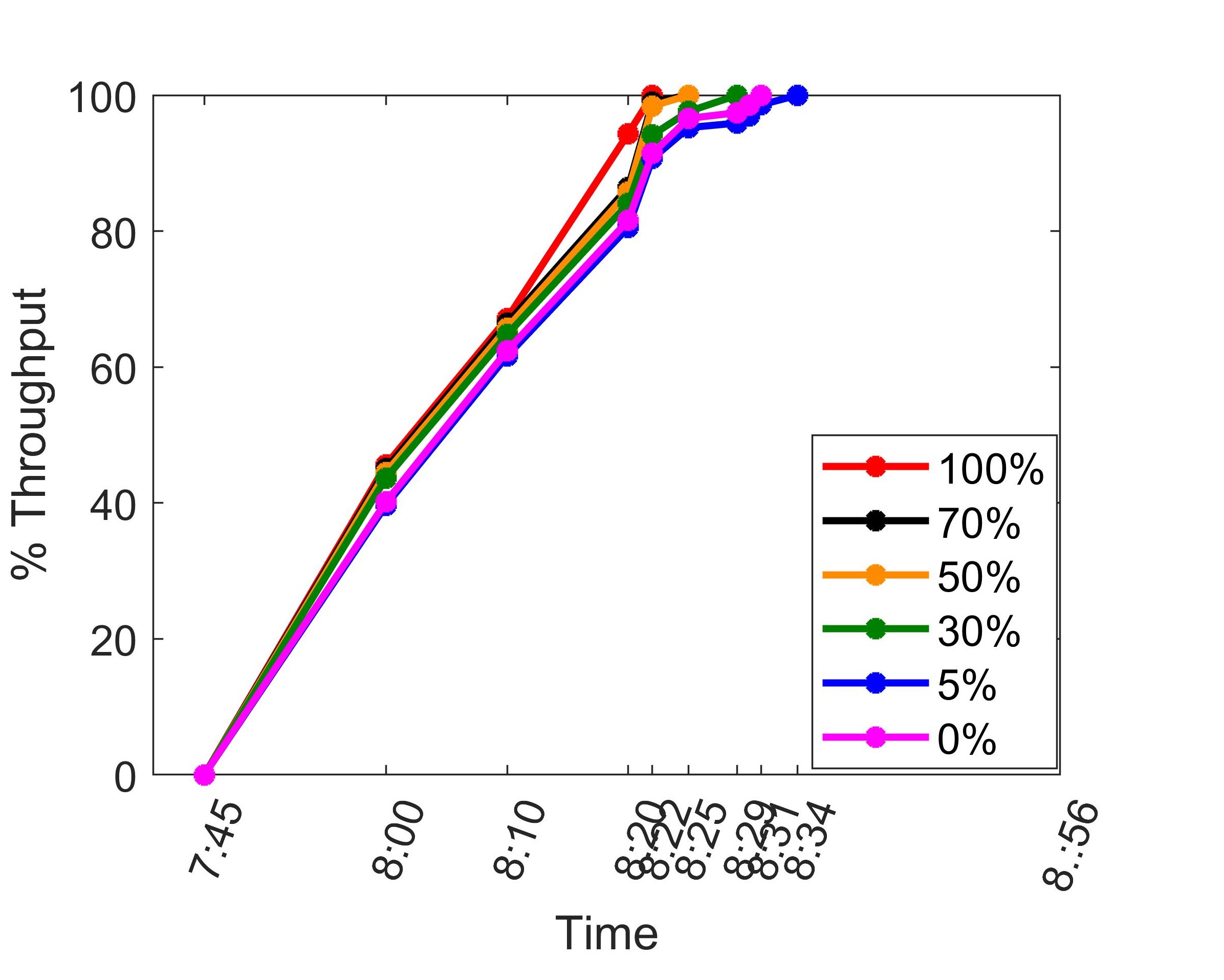}
	\caption{Throughput of Congested Network for Different MPRs}
	\label{Throughput_C_Impact_Paper}
\end{figure}

\begin{figure}[!t]
	\centering
	\includegraphics[width=3in]{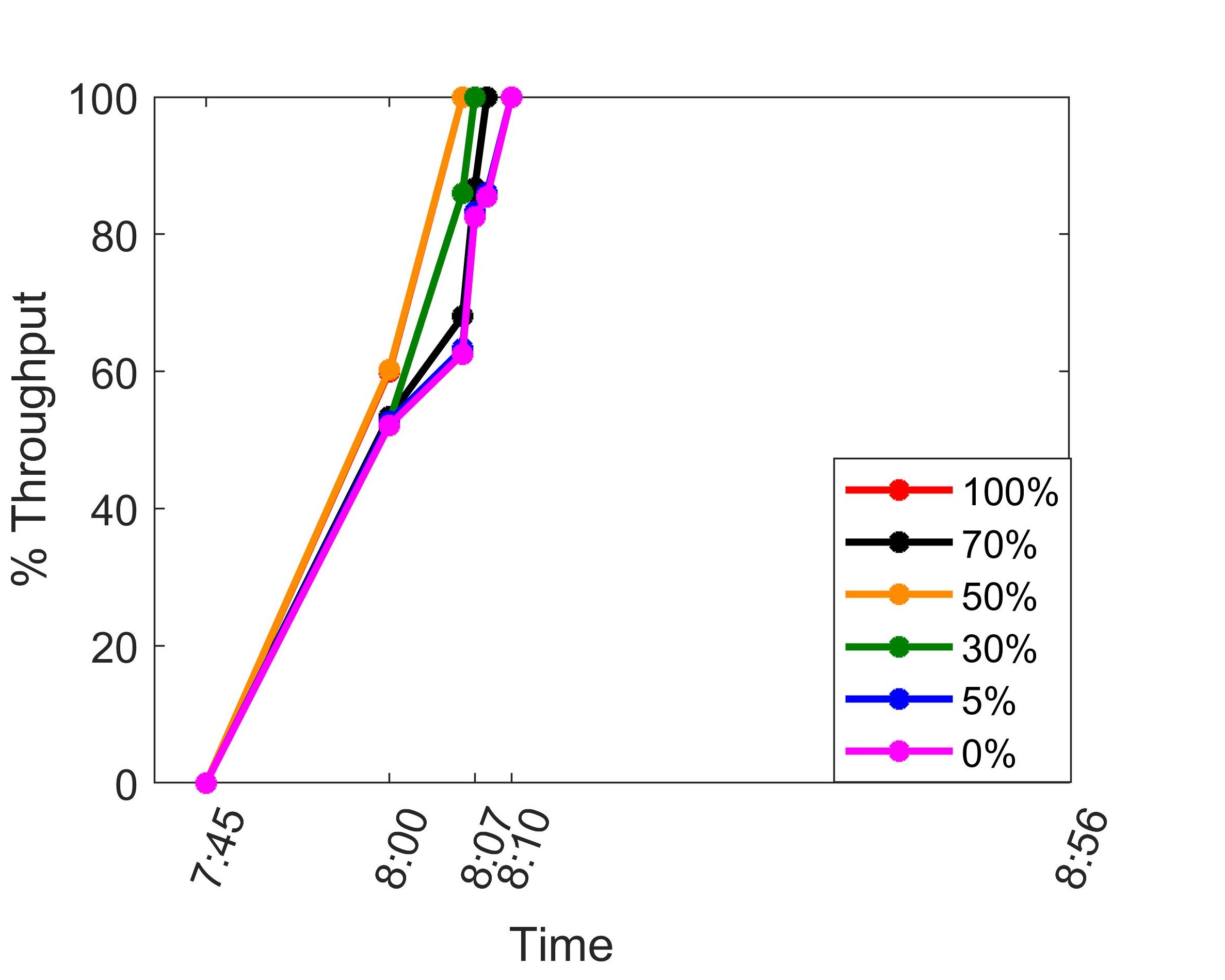}
	\caption{Throughput of Uncongested Network for Different MPRs}
	\label{Throughput_UC_Impact_Paper}
\end{figure}

Fig. \ref{Throughput_HC_Impact_paper}, Fig. \ref{Throughput_C_Impact_Paper}, and Fig. \ref{Throughput_UC_Impact_Paper} represent respectively the throughput for the three traffic conditions considered and show that the higher the MPR of CAVs the better the throughput and the less time required to achieve 100\% throughput. In other words, the traffic network is loaded and unloaded faster with higher MPRs of CAVs. In addition, it is important to note that for the uncongested traffic condition, throughput is reached only with a negligible time difference based on the MPR of CAVs employed.


\subsubsection{Average Speed}
\begin{figure*}[!ht]
	\centering
	\includegraphics[width=6in]{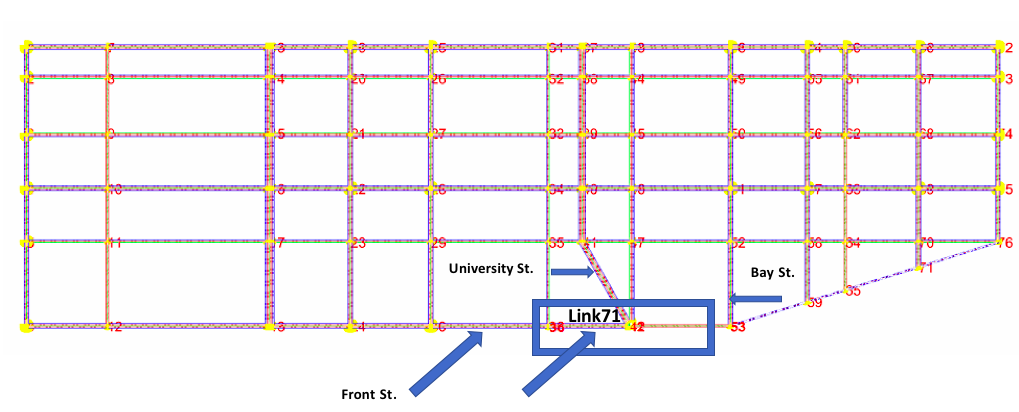}
	\caption{Downtown Toronto and the Most Congested Link}
	\label{fig11}
\end{figure*}

\begin{figure}[!th]
	\centering
	\includegraphics[width=3in]{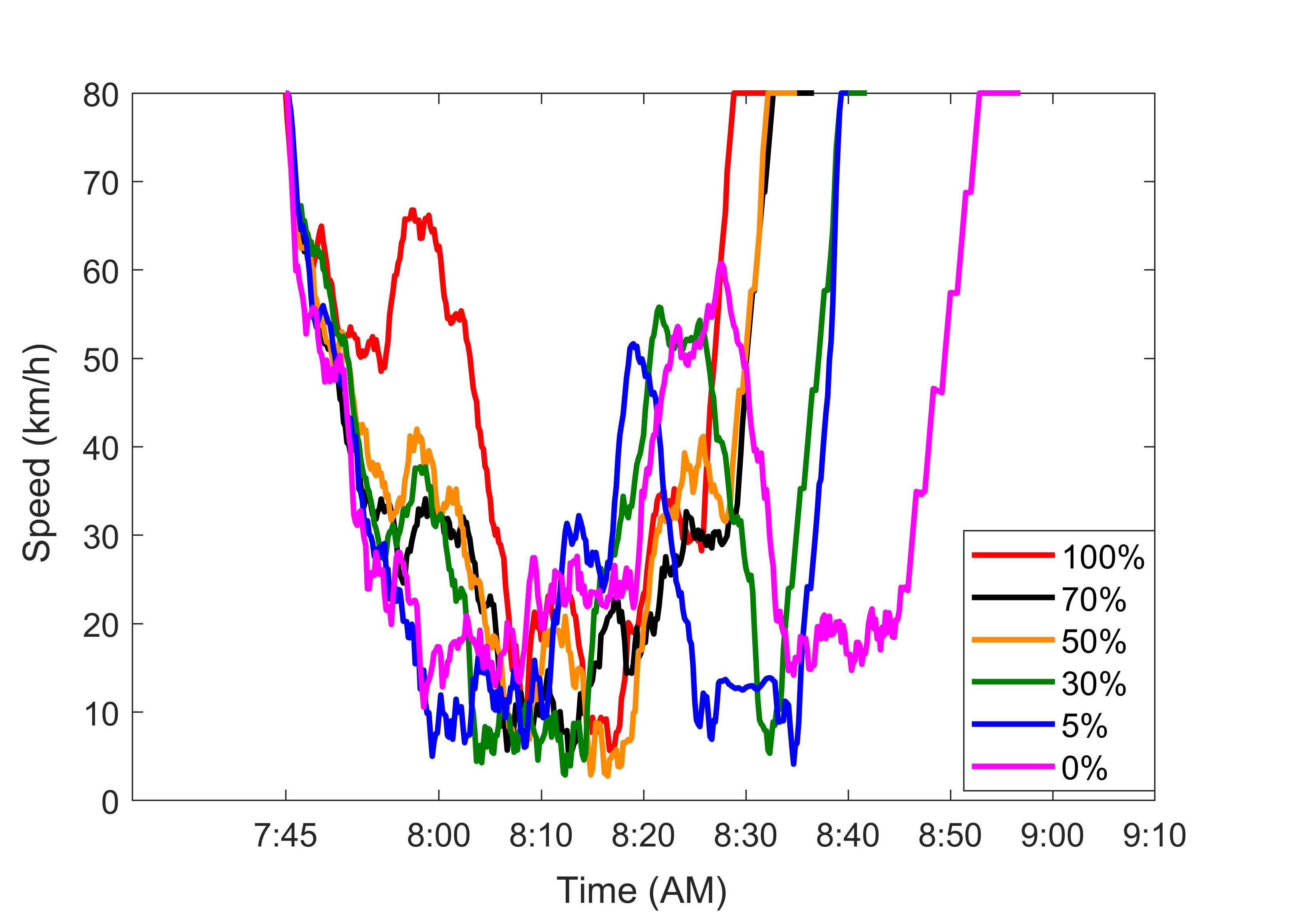}
	\caption{Average Link Speed for Link 71 Versus Time}
	\label{Speed_Time_HC_Impact_Paper}
\end{figure}

Average speed is a profound indicator that reflects the performance of a traffic network. The highly congested condition is considered to show the impact of the MPRs of CAVs in downtown Toronto over time. We first defined the most congested link as in Fig. \ref{fig11} in the network and that is of two lanes on Front Street and of 80km/h speed, which is link 71. The downstream link is a one-way lane link with 40km/h speed limit. In order to reduce the local noise of the data points, data smoothing was conducted. Fig. \ref{Speed_Time_HC_Impact_Paper} demonstrates that employing CAVs improves the average speed on the most congested link in the network. For the most congested link, the improvement of average speed when employing 100\% CAVs reaches as high as 50km/h compared to the case of 5\% CAVs. Higher average speed means less congestion as well.

\subsubsection{Average Density}
\begin{figure}[!t]
	\centering
	\includegraphics[width=3in]{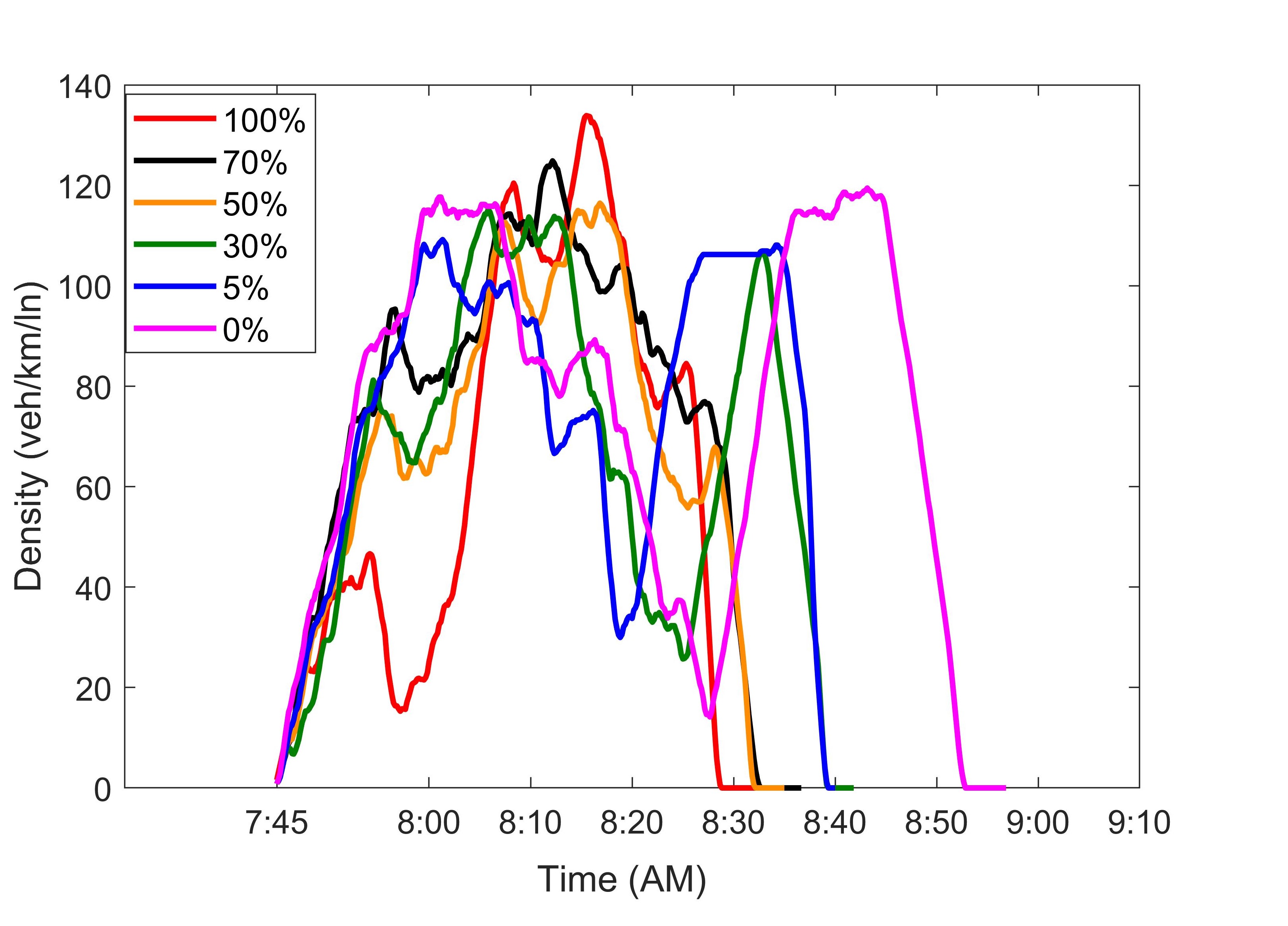}
	\caption{Average Link Density for Link 71 Versus Time}
	\label{Density_Time_HC_Impact_Paper}
\end{figure}

Reducing the density over time means that traffic networks face less congestion. In order to reduce the local noise of the data points, data smoothing was conducted. Fig. \ref{Density_Time_HC_Impact_Paper} shows that for 0\% CAVs and for the periods 8:00-8:10AM and 8:35-8:45AM, the density is close to the jam density. On the other hand, when employing higher MPRs of CAVs, the network is unloaded faster. Fig. \ref{Density_Time_HC_Impact_Paper} illustrates that for 100\% CAVs the link is unloaded around half an hour earlier compared to the 0\% CAVs. The main justification is that CAVs are being routed based on the up to date information provided by the intelligent intersections. Hence, CAVs are able to re-route and en-route to avoid gridlocks. It is crucial to note as well that since CAVs have lower safe spacing distance and reaction time, a slight increase in density is observed during the peak period. This means that the link can accommodate more vehicles.

\section{Conclusion}
In this study we demonstrated the effectiveness of a distributed routing system for intelligent vehicles (CAVs) at various levels of MPRs and traffic conditions on a large-scale urban network of downtown Toronto. We adopted the end-to-end distributed routing system (E2ECAV) proposed by Djavadian and Farooq \cite{r7} and started with defining the optimal set of parameters needed for vehicular movement model (IDM) and E2ECAV. It was found in our sensitivity analysis that 60sec is the optimal and realistic updating interval after taking into account the communication and processing delays.  Although it has been reported that the reduced safe spacing distance and reaction time for the IDM implementation have substantial impact on highway traffic, in urban traffic networks their impact is only slightly better and can be neglected. However, this triggers the need for connectivity as it has been shown in our study that with the distributed routing, the higher the MPR of CAVs the better the traffic network characteristics. For the highly congested case, 100\% CAVs resulted in an increase in average speed of as high as 50km/h for the most congested link. Furthermore, it is clear from the density over time figure of the most congested link and for the highly congested traffic condition that the higher the MPRs of CAVs the faster the network is unloaded. Moreover, less travel time, and better throughput. For the highly congested traffic network, employing 50\%, 70\%, and 100\% has relatively similar amount of reduction of the mean travel time. Around 18\% reduction of the mean travel time is observed when employing 100\% CAVs. Furthermore, for 50\%, 70\%, and 100\% VACs, a slight increase of the mean vehicle kilometers travelled and of 4\% when employing 100\% CAVs. It is crucial to note that the impact of higher MPRs of CAVs is profound in the case of congested and highly congested traffic networks.
In future studies we argue more emphasis on the effect of CAVs and on setting strategies that prevent looping and rerouting for more efficient algorithms that can be employed in the case of distributed routing systems. Furthermore, more replications of every scenario would give more consistent results and show the trend more accurately. Finally, investigating the impact of the proposed E2ECAV routing system environmentally would be a good suggestion for further future researches. 


%

\pagebreak
\section*{Acknowledgment}
This research is funded by Ryerson University, Ontario Early Researcher Award, and NSERC Canada Research Chair on Disruptive Transportation Technologies and Services.



%


\end{document}